\documentclass[final,5p,times,twocolumn,authoryear]{elsarticle}

\usepackage{graphicx}
\usepackage{amssymb}
\usepackage{natbib}

\journal{Icarus}

\begin{document}

\begin{frontmatter}

\title{The Eos family halo}

\author{M.~Bro\v z}

\address{Institute of Astronomy, Charles University, Prague, V Hole\v sovi\v ck\'ach 2, 18000 Prague 8, Czech Republic e-mail: mira@sirrah.troja.mff.cuni.cz}

\author{A.~Morbidelli}

\address{Observatoire de la C\^ote d'Azur, BP 4229, 06304 Nice Cedex 4, France, e-mail: morby@oca.eu}

%%%%%%%%%%%%%%%%%%%%%%%%%%%%%%%%%%%%%%%%%%%%%%%%%%%%%%%%%%%%%%%%%%%%%%%%

\begin{abstract}
We study K-type asteroids in the broad surroundings of the Eos family
because they seem to be intimately related, according to their colours measured by the Sloan Digital Sky Survey.
Such `halos' of asteroid families have been rarely used as constraints
for dynamical studies to date. We explain its origin as bodies escaping
from the family `core' due to the Yarkovsky semimajor-axis drift and interactions
with gravitational resonances, mostly with the 9/4 mean-motion resonance with Jupiter at 3.03\,AU.
Our $N$-body dynamical model allows us to independently estimate the age of the family $1.5\hbox{ to }1.9$\,Gyr.
This is approximately in agreement with the previous age estimate by \citet{Vokrouhlicky_etal_2006Icar..182...92V}
based on a simplified model (which accounts only for changes of semimajor axis).
We can also constrain the geometry of the disruption event
which had to occur at the true anomaly $f \simeq 120^\circ\hbox{ to }180^\circ$.
\end{abstract}


\end{frontmatter}

% \linenumbers

%%%%%%%%%%%%%%%%%%%%%%%%%%%%%%%%%%%%%%%%%%%%%%%%%%%%%%%%%%%%%%%%%%%%%%%%

\section{Introduction}\label{sec:introduction}

The Eos family is one of the best-studied families in the main asteroid belt.
Although we do not attempt to repeat a thorough review presented
in our previous paper \citet{Vokrouhlicky_etal_2006Icar..182...92V},
we recall that the basic structure of the family is the following:
  (i)~there is a sharp inner boundary coinciding with the 7/3 mean-motion resonance with Jupiter
      at approximately 2.96\,AU;
 (ii)~the 9/4 mean-motion resonance with Jupiter divides the family at 3.03\,AU
      and asteroids with larger sizes are less numerous at larger semimajor axes;
(iii)~there is an extension of the family along the $z_1 \equiv g - g_6 + s - s_6$
      secular resonance towards lower values of proper semimajor axis $a_{\rm p}$,
      eccentricity~$e_{\rm p}$ and inclination $\sin I_{\rm p}$.
All these fact seem to be determined by the interaction between the orbits
drifting due to the Yarkovsky effect in semimajor axis and
the gravitational resonances which may affect eccentricities and inclinations.

In this work, we focus on a 'halo' of asteroids around the nominal Eos family
which is clearly visible in the Sloan Digital Sky Survey, Moving Object Catalogue version~4
(SDSS, \citealt{Parker_etal_2008Icar..198..138P}). As we shall see below, both the `halo'
and the family have the same SDSS colours and are thus most likely related to each other.
Luckily, the Eos family seems to be spectrally distinct in this part of the main belt
(several Eos family members were classified as K-types by \citealt{DeMeo_etal_2009Icar..202..160D})
and it falls in between S-complex and C/X-complex asteroids in terms of the SDSS colour indices.
Detailed spectroscopic observations were also performed by \cite{Zappala_etal_2000Icar..145....4Z}
which confirmed that asteroids are escaping from the Eos family
due to the interaction with the J9/4 resonance.

Our main motivation is to understand the origin of the whole halo
and to explain its unusually large spread in eccentricity and inclination
which is hard to reconcile with any reasonable initial velocity field.
Essentially, this is a substantial extension of work of \citet{Vokrouhlicky_etal_2006Icar..182...92V},
but here we are interested in bodies which {\em escaped\/} from the nominal family.%

We were also curious if such halos may be somehow related
to the giant-planet migration which would have caused significant gravitational perturbations
of all small-body populations \citep{Morbidelli_etal_2005Natur.435..462M}.
Of course, in such a case the process is {\em size-independent\/}
and moreover the age of the corresponding family would have to approach 3.9\,Gyr
in order to match the Nice model of giant-planet migration.

In Section~\ref{sec:halo_core}, we define the Eos halo and core populations.
Section~\ref{sec:yarko} is devoted to a description of our dynamical model
and to a comparison with the SDSS observations.
We discuss consequences of our results in Section~\ref{sec:conclusions}.

%%%%%%%%%%%%%%%%%%%%%%%%%%%%%%%%%%%%%%%%%%%%%%%%%%%%%%%%%%%%%%%%%%%%%%%%

\section{A discernment of the family core and halo}\label{sec:halo_core}

In this Section, we proceed as follows:
  (i)~we use a hierarchical clustering method to extract the nominal Eos family;
 (ii)~we look at the members of the family with SDSS colours and we define a colour range;
(iii)~we select all asteroids with Eos-like colours from the SDSS catalogue; finally,
 (iv)~we define a halo and core using simple 'boxes' in the proper-element space.

\subsection{Colours of Eos-like asteroids}

\begin{figure}
\centering
\includegraphics[width=5cm]{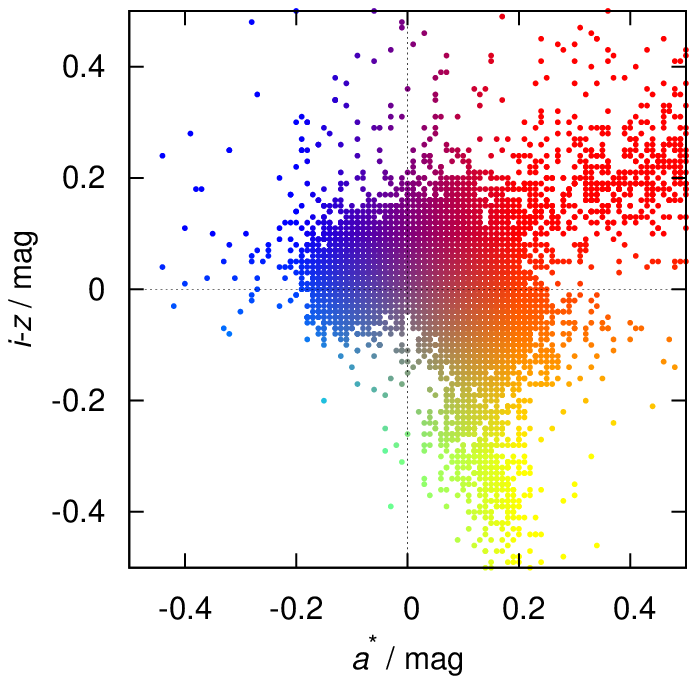}
\includegraphics[width=5cm]{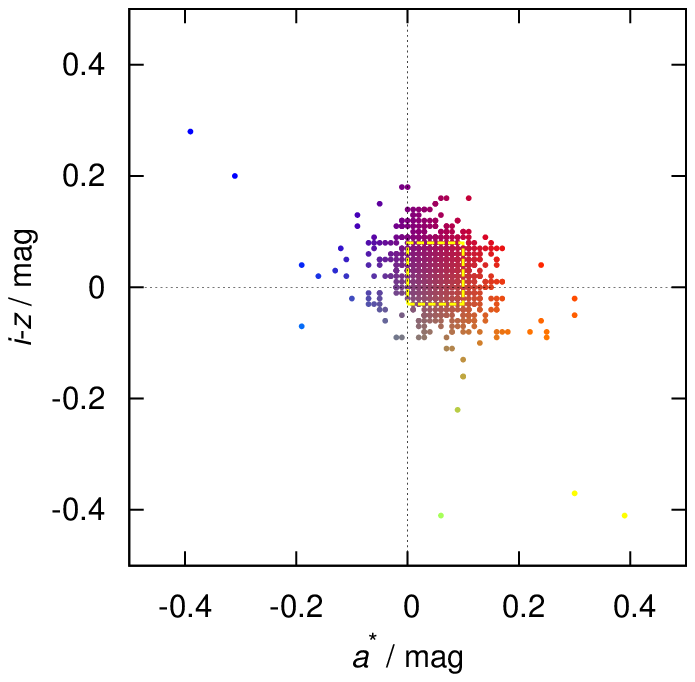}
\caption{Colour indices $i-z$ and $a^*$ (defined in \citet{Parker_etal_2008Icar..198..138P})
of all asteroids from the Sloan Digital Sky Survey, Moving Object Catalogue version~4
and the corresponding colour palette (top panel)
which is used in the following figures to distinguish colours of asteroids.
We also plot the Eos family members observed by the SDSS (bottom panel)
with small photometric uncertainties (less than 0.03\,mag).
The inferred range of colour indices (denoted by the dashed yellow rectangle)
is then used as a criterion for the selection of the Eos-like asteroids
in the broad surroundings of the nominal family.
The rectangle does not encompass the outliers.}
\label{a_star_i-z_COLOR_ALL_SMALLERR}
\end{figure}

We want to select asteroids similar to the Eos family,
but first we have to choose a criterion to do so.
We thus identify the nominal Eos family using
a hierarchical clustering method (HCM, \citealt{Zappala_etal_1995Icar..116..291Z})
with a suitably low cut-off velocity $v_{\rm cutoff} = 50\,{\rm m}/{\rm s}$
(which leads to a similar extent of the family as in \cite{Vokrouhlicky_etal_2006Icar..182...92V}),
and extract colour data from the SDSS catalogue (see Figure~\ref{a_star_i-z_COLOR_ALL_SMALLERR}).
The majority of Eos-family asteroids have colour indices in the following intervals
\begin{eqnarray}
a^* & \in & ( 0.0 , 0.1 )\,{\rm mag}\,, \\
i-z & \in & (-0.03, 0.08)\,{\rm mag}
\end{eqnarray}
which then serves as a criterion for the selection of Eos-like asteroids
in the broad surroundings of the nominal family.

We also used an independent method for the selection of Eos-like asteroids
employing a 1-dimensional colour index (which was used in \citealt{Parker_etal_2008Icar..198..138P}
to construct their colour palette)
and we verified that our results are not sensitive to this procedure.

%%%%%%%%%%%%%%%%%%%%%%%%%%%%%%%%%%%%%%%%%%%%%%%%%%%%%%%%%%%%%%%%%%%%%%%%

\subsection{Boundaries in the proper element space}

\begin{figure}
\centering
\includegraphics[width=6cm]{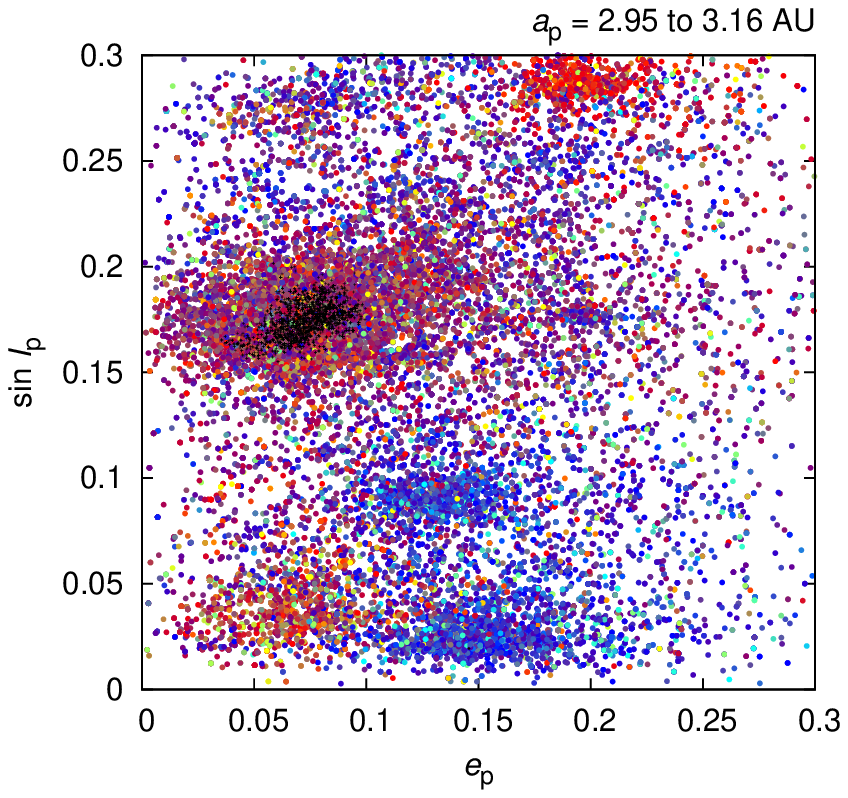}
\includegraphics[width=6cm]{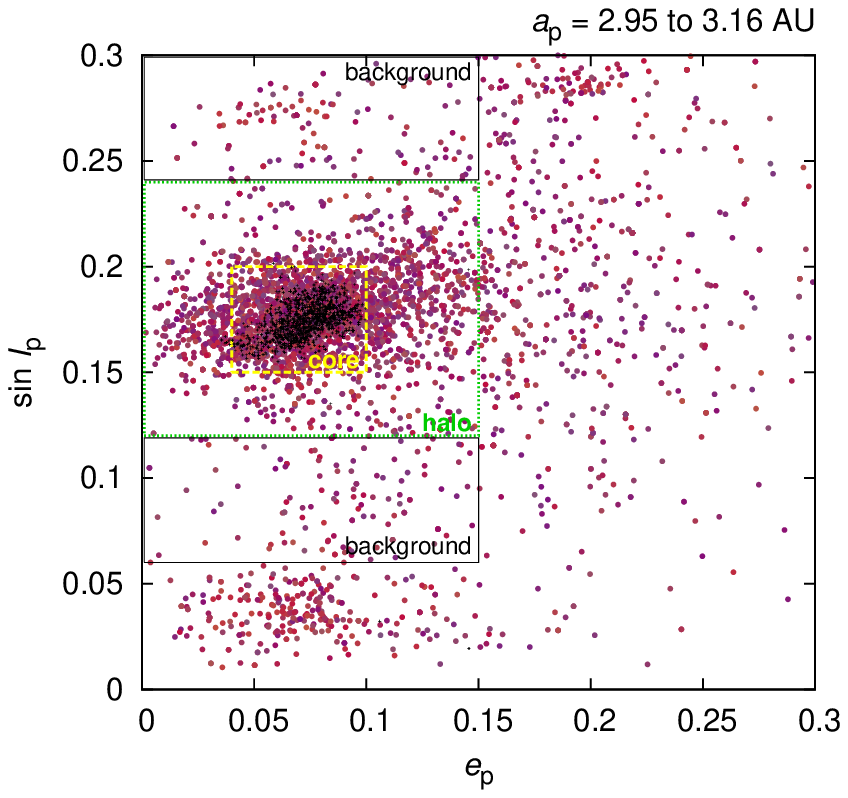}
\caption{The proper eccentricity~$e_{\rm p}$ vs proper inclination~$\sin I_{\rm p}$ plot
for asteroids included in the SDSS MOC~4 catalogue.
The proper semimajor axis is confined to the interval 2.95~to 3.16\,AU, i.e. the Eos family zone.
Colour coding corresponds to the SDSS colour indices according to Figure~\ref{a_star_i-z_COLOR_ALL_SMALLERR}.
The top panel includes {\em all\/} asteroids (regardless of their colours).
The bottom panel shows only a subset of 'Eos-like' asteroids with colours similar
to those of the Eos members (see Fig.~\ref{a_star_i-z_COLOR_ALL_SMALLERR}, bottom).
Moreover, we denote a box used for the definition of the family 'core' (dashed yellow line)
a larger box for the 'halo' (dotted green line)
and two boxes considered as 'background' (thin black line).
For comparison, we also plot positions of the nominal Eos family members (black dots),
identified for the velocity $v_{\rm cutoff} = 50\,{\rm m}/{\rm s}$.}
\label{eos_halo_morby_BOX_ei_COLOR}
\end{figure}

\begin{figure}
\centering
\includegraphics[height=5.5cm]{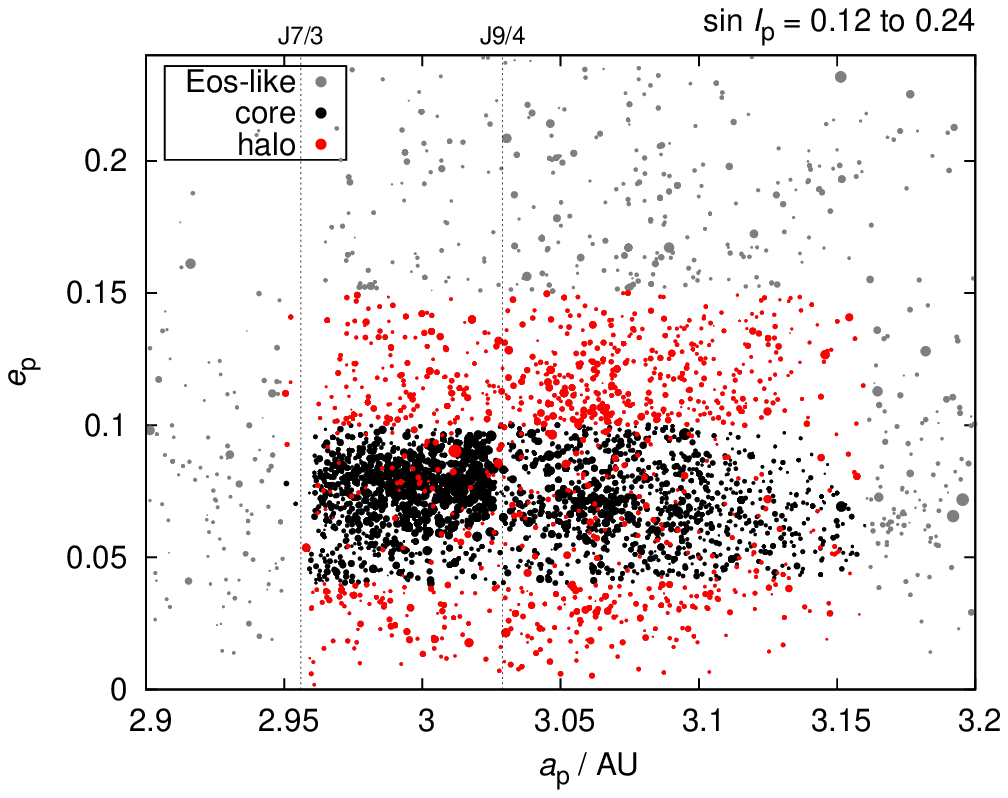}
\includegraphics[height=5.5cm]{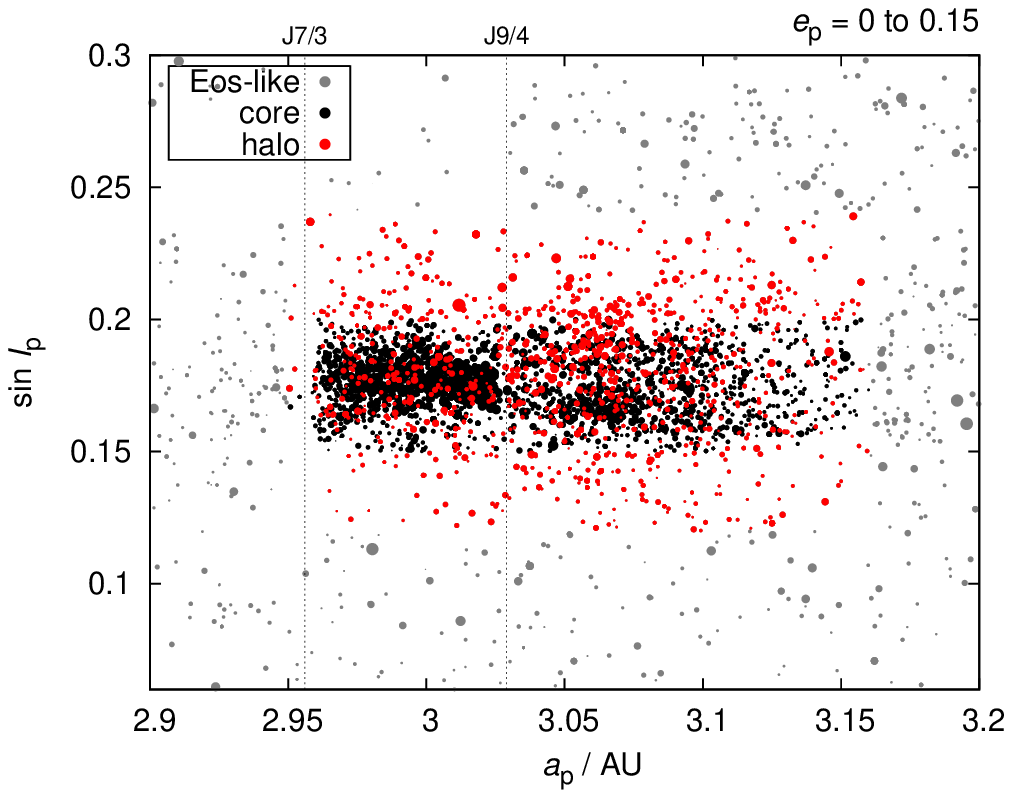}
\caption{The proper semimajor axis~$a_{\rm p}$ vs proper eccentricity~$e_{\rm p}$ (top panel)
and $a_{\rm p}$ vs proper inclination~$\sin I_{\rm p}$ (bottom panel)
for the observed Eos core (black dots), halo (red dots) and remaining Eos-like asteroids (gray dots) in the surroundings.
The sizes of symbols are (inversely) proportional to the absolute magnitudes~$H$ of asteroids.
The positions of important mean-motion resonances with Jupiter are also indicated (dotted vertical lines).}
\label{eos_halo_morby_BOX_ae_SIZES}
\end{figure}

Next, we have to distinguish the family 'core' and 'halo' populations
on the basis of proper orbital elements $(a_{\rm p}, e_{\rm p}, \sin I_{\rm p})$
which will be consistently used for both the SDSS observations and our dynamical models.
We also need to define 'background' population which enables
to estimate how many asteroids might have Eos-like colours by chance.
We decided to use a simple box criterion (see Figures~\ref{eos_halo_morby_BOX_ei_COLOR},
\ref{eos_halo_morby_BOX_ae_SIZES}
and Table~\ref{core_halo_background}),
while the range of proper semimajor axis is always the same,
$a_{\rm p} \in (2.95, 3.16)\,{\rm AU}$.

\begin{table}
\caption{The definitions of the core, halo and background populations
in terms of intervals of proper eccentricity~$e_{\rm p}$ and proper inclination~$\sin I_{\rm p}$.
The range of proper semimajor axis $a_{\rm p} \in (2.95, 3.16)\,{\rm AU}$ is the same
in all cases.}
\label{core_halo_background}
\centering
\medskip
\begin{tabular}{rccl}
\hline
population & $e_{\rm p}$ & $\sin I_{\rm p}$ & note \\
\hline
core       & 0.04--0.10 & 0.15--0.20 \\
halo       & 0.00--0.15 & 0.12--0.24 & and {\em not\/} in the core \\
background & 0.00--0.15 & 0.06--0.12 & together with\dots \\
           & 0.00--0.15 & 0.24--0.30 \\ 
\hline
\end{tabular}
\end{table}

Our results do not depend strongly on the selection criterion.
For example, we tested a stringent definition:
core was identified by the HCM at $v_{\rm cutoff} = 50\,{\rm m}/{\rm s}$
and all remaining bodies in the surroundings belong to the halo.
This approach makes the core as small as possible and the halo correspondingly larger
but our results below (based on halo/core ratios) would be essentially the same.
According to our tests, not even a different definition of the background/halo boundary
changes our results.

\begin{figure}
\centering
\includegraphics[width=6.5cm]{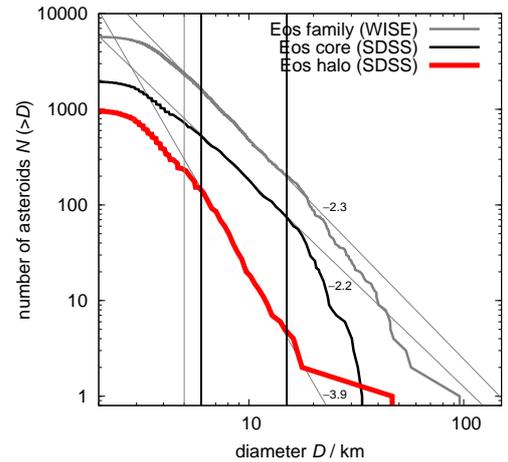}
\caption{The cumulative size-frequency distributions $N({>}D)$ of the Eos core and halo.
We show power-law fits and corresponding slopes~$\gamma$ which clearly indicate
that the halo population is significantly {\em steeper\/} than the core population.
For comparison, we also plot the SFD of the nominal Eos family (as inferred from the WISE data, \citealt{Masiero_etal_2011ApJ...741...68M}).
The SFD's of the core and halo are biased because they include
{\em only\/} asteroids observed by the SDSS. Consequently, the core seems to be much less
populated than the nominal Eos family, even though these SFD's should be very similar.
Nevertheless, the slopes and the halo/core ratios which we use in our analysis is not
much affected by this bias.}
\label{eos_halo_morby_BOX_size_distribution_SMALLD_WISE}
\end{figure}

We are now ready to construct size-frequency distributions of individual populations.
In order to convert absolute magnitudes~$H$ to diameters~$D$
we computed the median geometric albedo $p_V = 0.16$
from the WISE data \citep{Masiero_etal_2011ApJ...741...68M} for the nominal Eos family members.
The size-frequency distribution (Figure~\ref{eos_halo_morby_BOX_size_distribution_SMALLD_WISE}) of the halo
has a cumulative slope $N(>D) \propto D^\gamma$ equal to $\gamma = -3.9 \pm 0.2$
in the size range $D = 6\hbox{ to }15\,{\rm km}$
and is significantly {\em steeper\/} than that of the core ($\gamma = -2.2 \pm 0.1$).
Even this difference of slopes $(1.7 \pm 0.2)$ indicates that if there
a process transporting asteroids from the core to the halo it must be indeed size-dependent.

A~frequency analysis similar as in \cite{Carruba_Michtchenko_2007A&A...475.1145C} or \cite{Carruba_2009MNRAS.395..358C}
shows that there is approximately 5\,\% of likely $z_1$ resonators
(with the frequency $g-g_6 + s-s_6 < 0.3\,''/{\rm yr}$) in the halo region.
However, the concentration of objects inside and outside
the resonance is roughly the same, so that this secular resonance
does not seem to be the most important transport mechanism.

%%%%%%%%%%%%%%%%%%%%%%%%%%%%%%%%%%%%%%%%%%%%%%%%%%%%%%%%%%%%%%%%%%%%%%%%

\section{Yarkovsky-driven origin of the halo}\label{sec:yarko}

Motivated by the differences of the observed SFD's,
we now want to test a hypothesis that the Eos family halo
(or at least a part of it) was created by the Yarkovsky
semimajor-axis drift, which pushes objects from the core
into neighbouring mean-motion resonances and consequently
to the halo region.

\subsection{Initial conditions}

We prepared an $N$-body simulation of the long-term evolution of the Eos core and halo
with the following initial conditions:
we included the Sun and the four giant planets on current orbits.
We applied a standard barycentric correction to both massive objects
and test particles to prevent a substantial shift of secular frequencies \citep{Milani_Knezevic_1992Icar...98..211M}.
The total number of test particles was 6545,
with sizes ranging from $D = 104\hbox{ to }1.5\,{\rm km}$
and the distribution resembling the observed SFD of the Eos family.

Material properties were as follows:
the bulk density $\rho = 2500\,{\rm kg}/{\rm m}^3$,
the surface density $\rho_{\rm s} = 1500\,{\rm kg}/{\rm m}^3$,
the thermal conductivity $K = 0.001\,{\rm W}/{\rm m}/{\rm K}$,
the specific thermal capacity $C = 680\,{\rm J}/{\rm kg}/{\rm K}$,
the Bond albedo $A = 0.1$,
the infrared emissivity $\epsilon = 0.9$,
i.e. all typical values for regolith covered basaltic asteroids.

Initial rotation periods were distributed uniformly on the interval 2~to 10\,hours
and we used random (isotropic) orientations of the spin axes.
The YORP model of the spin evolution was described in detail in \citet{Broz_etal_2011MNRAS.414.2716B},
while the efficiency parameter was $c_{\rm YORP} = 0.33$
(i.e. a likely value according to \citealt{Hanus_etal_2011A&A...530A.134H}).
YORP angular momenta affecting the spin rate and the obliquity were taken from \citet{Capek_Vokrouhlicky_2004Icar..172..526C}.
We also included spin axis reorientations caused by collisions%
\footnote{We do {\em not} take into account collisional disruptions
because we model only that subset of asteroids which
survived subsequent collisional grinding (and compare
it to the currently observed asteroids). Of course,
if we would like to discuss e.g. the size of the parent body,
it would be necessary to model disruptive collisions too.}
with a time scale estimated by \citet{Farinella_etal_1998Icar..132..378F}:
$\tau_{\rm reor} = B \left({\omega/\omega_0}\right)^{\beta_1} \left({D/D_0}\right)^{\beta_2}$,
where
$B = 84.5\,{\rm kyr}$,
$\beta_1 = 5/6$,
$\beta_2 = 4/3$,
$D_0 = 2\,{\rm m}$ and
$\omega_0$ corresponds to period $P = 5$~hours.

The initial velocity field was size-dependent,
$v \propto v_0 D_0/D$,
with $v_0 = 93\,{\rm m}/{\rm s}$ and $D_0 = 5\,{\rm km}$ (i.e. the best-fit values from \citealt{Vokrouhlicky_etal_2006Icar..182...92V}).
In principle, this type of size--velocity relation was initially
suggested by \cite{Cellino_etal_1999Icar..141...79C},
but here, we attempt to interpret the structure of the family
as a complex interplay between the velocity field and the Yarkovsky drift
which is also inversely proportional to size.
We assumed isotropic orientations of the velocity vectors.
The geometry of collisional disruption was determined by
the true anomaly $f = 150^\circ$,
and the argument of perihelion $\omega = 30^\circ$.
We discuss different geometries in Section~\ref{sec:conclusions}.

We use a modified version of the SWIFT package \citep{Levison_Duncan_1994Icar..108...18L}
for numerical integrations,
with a second-order symplectic scheme \citep{Laskar_Robutel_2001CeMDA..80...39L},
digital filters employing frequency-modified Fourier transform \citep{Sidlichovsky_Nesvorny_1996CeMDA..65..137S}
and an implementation of the Yarkovsky effect \citep{Broz_2006}.
The integration time step was~$\Delta t = 91\,{\rm days}$,
the output time step after all filtering procedures~$10\,{\rm Myr}$
and the total integration time span reached~4\,Gyr.

%%%%%%%%%%%%%%%%%%%%%%%%%%%%%%%%%%%%%%%%%%%%%%%%%%%%%%%%%%%%%%%%%%%%%%%%

\subsection{Results of the $N$-body simulation}

\begin{figure*}
\centering
\begin{tabular}{cc}
\includegraphics[height=5.0cm]{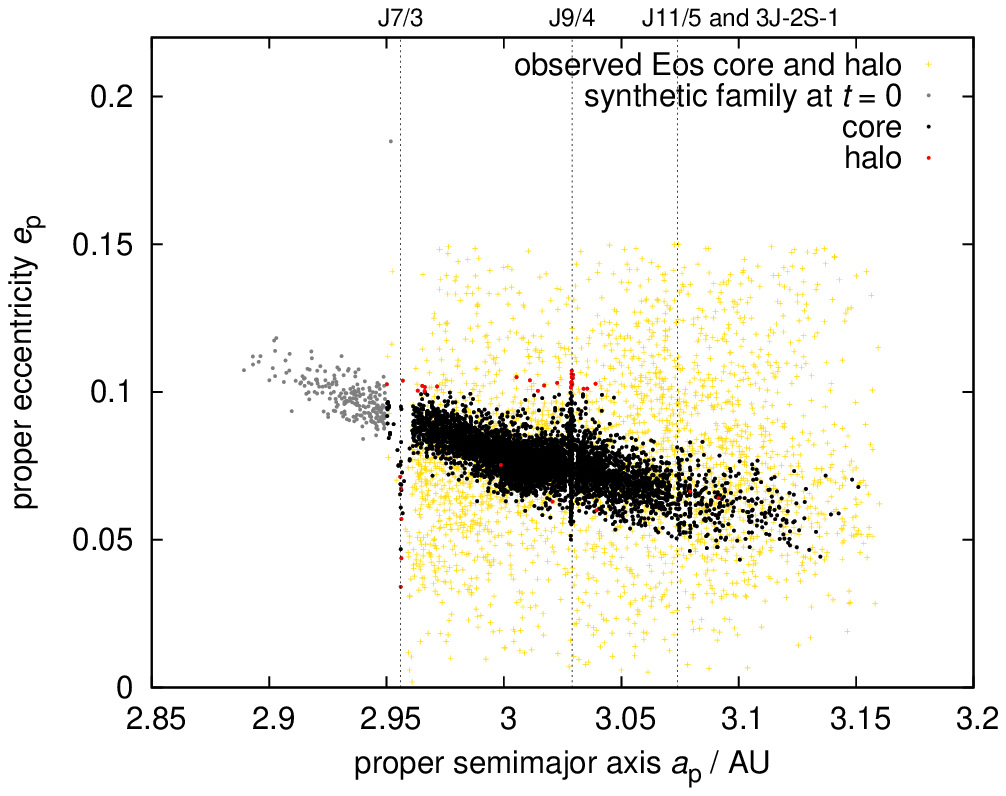} &
\includegraphics[height=4.9cm]{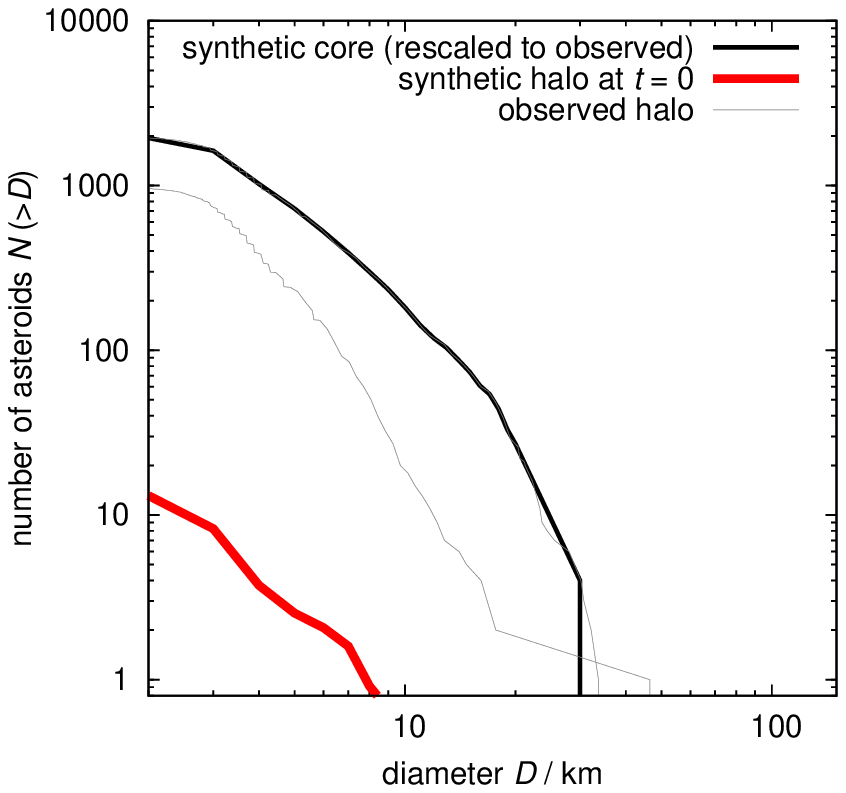} \\
\includegraphics[height=5.0cm]{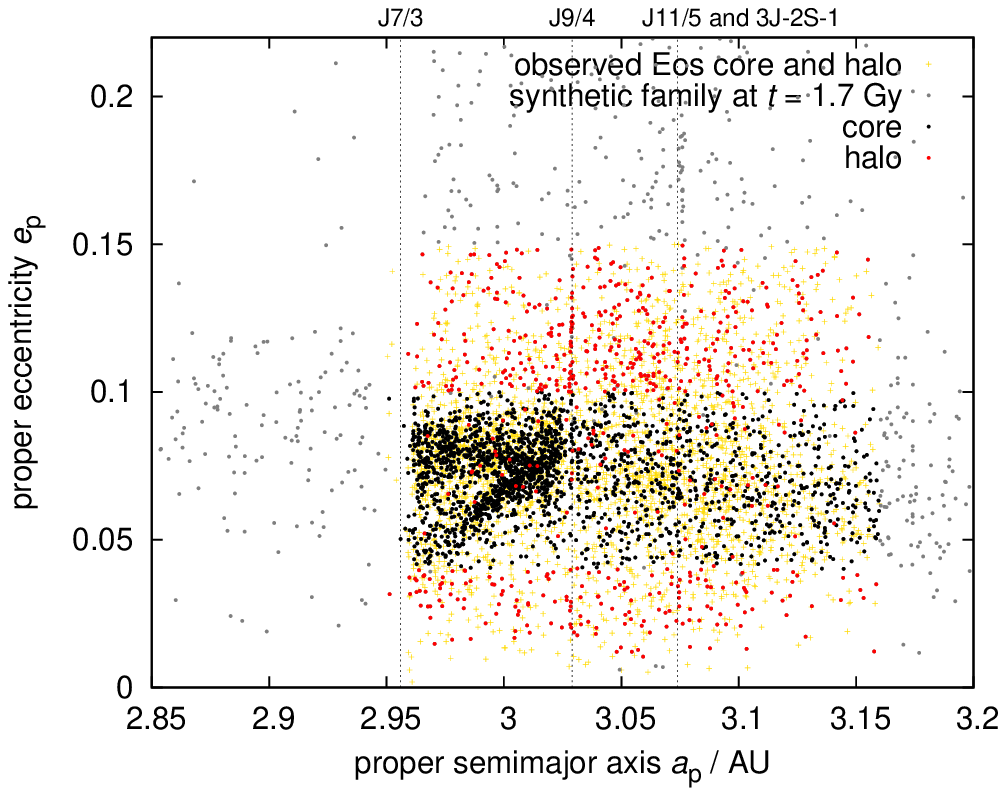} &
\includegraphics[height=4.9cm]{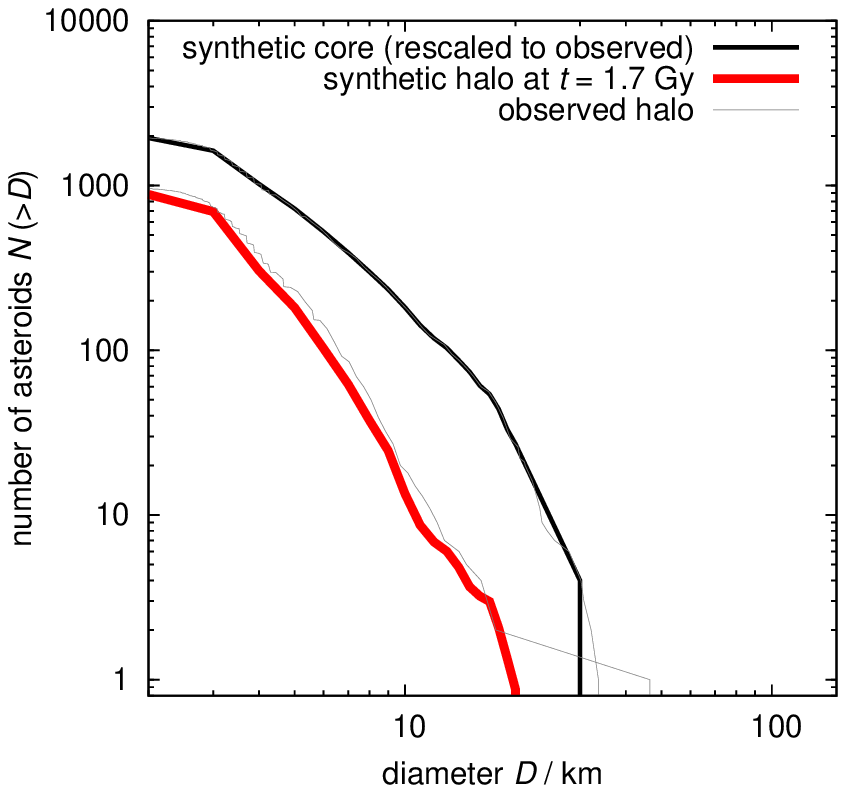} \\
\end{tabular}
\caption{Left panels: the proper semimajor axis~$a_{\rm p}$ vs proper eccentricity~$e_{\rm p}$ plots
showing a dynamical evolution of our synthetic family. We can distinguish
the core (black dots),
the halo (red dots) and
objects beyond the halo box (gray dots).
There is a comparison to the observed Eos core and halo too (yellow crosses),
as inferred from the SDSS data (the same as in Figure~\ref{eos_halo_morby_BOX_ei_COLOR}, bottom).
The positions of important resonances are indicated by vertical dotted lines.
We plot the initial situation at $t = 0$ (top panel)
and the evolved family at $t = 1.7$\,Gyr (bottom panel).
The core of the synthetic family exhibits a slightly different
structure than the observed core which may indicate that:
(i)~the initial true anomaly was closer to $f = 180^\circ$, or
(ii)~the initial velocity field deviated from the assumed $v \propto 1/D$ dependence.
Right panels: the corresponding size-frequency distributions of the synthetic core (black line),
which was always scaled to the observed SFD of the Eos core, and the synthetic halo (red line)
which can be then directly compared to the observed halo (gray line).
It is clear that the halo's SFD becomes steeper in the course of time and at $t \simeq 1.7$\,Gyr
it matches the observed SFD.}
\label{eos-5_halo_core_ae_0000My}
\end{figure*}

Initially, almost all asteroids are located in the core (see Figure~\ref{eos-5_halo_core_ae_0000My}).
Only a few outliers may have velocities large enough to belong to the halo.
Within a few million years the halo/core ratio quickly increases due to objects
located inside the 9/4 resonance and injected to the halo by these
size-independent gravitational perturbations.
Further increase is caused by the Yarkovsky/YORP semimajor axis drift
which pushes additional orbits into the J9/4 and also other resonances.

\begin{figure}
\centering
\includegraphics[width=8cm]{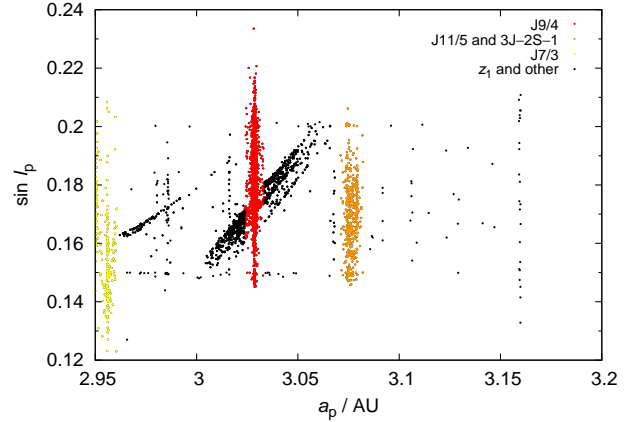}
\caption{The proper semimajor axis~$a_{\rm p}$ vs proper inclination~$\sin I_{\rm p}$
of the synthetic family members at the moment when they have entered the halo region.
It is easy to distinguish objects injected by mean-motion resonances
and by secular resonances in this projection, because the former have
a particular value of the semimajor axis. The objects injected by
the J9/4 resonance are denoted by red colour,
the J11/5 and $3{\rm J}-2{\rm S}-1$ by orange,
the J7/3 by yellow,
the $z_1$ secular resonance and other resonances by black.}
\label{eos-5_halo_injection_ai}
\end{figure}

We checked the orbital elements of bodies at the moment when they enter
the halo region (Figure~\ref{eos-5_halo_injection_ai}) and we computed
the statistics of dynamical routes that had injected bodies in the halo:
J9/4                                                                                        57\,\%,
J11/5 (together with a three-body resonance $3{\rm J}-2{\rm S}-1$ with Jupiter and Saturn)  10\,\%,
J7/3                                                                                         6\,\%, and
$z_1$~secular resonance                                                                     23\,\%.
The remaining few percent of bodies may enter the halo by different dynamical routes.%
\footnote{Other secular resonances intersecting this region, $s-s_6-2g_5+2g_6$
or $g+2g_5-3g_6$, do not seem to be important with respect to the transport
from the core to the halo.}
However, if we account for the fact that bodies captured by the $z_1$~resonance
usually encounter also the J9/4 resonance that scatters them further away
in to the halo, we obtain a modified statistics:
J9/4   70\,\%,
J11/5  12\,\%,
J7/3    5\,\%, and
$z_1$  10\,\%
that better reflects the importance of different mechanisms.

A saturation of the halo occurs after approximately $1\,{\rm Gyr}$,
because the halo population is affected by the Yarkovsky/YORP drift too,
so that the injection rate roughly matches the removal rate.
Nevertheless, the halo/core ratio steadily increases, which is
caused by the ongoing decay of the core population.

In order to compare our model and the SDSS observations we compute the ratio
${\cal R} = {{\rm d}N_{\rm halo}/{\rm d}N_{\rm core}}$
between the number of objects in the halo and in the core for a given {\em differential\/} size bin.
This can be computed straightforwardly from our simulation data.
In case of the SDSS observations, however, we think that there is a real {\em background\/}
of asteroids with Eos-like colours (may be due to observational uncertainties or a natural spread of colours;
see Figure~\ref{eos_halo_morby_BOX_ei_COLOR}).
Obviously, such background overlaps with the core and the halo,
so we need to subtract this contamination
\begin{equation}
{\cal R}_{\rm obs} \equiv { {\rm d}N_{\rm halo} - 0.833\, {\rm d}N_{\rm background} \over {\rm d}N_{\rm core} - 0.167\, {\rm d}N_{\rm background}}\,.\label{ratio}
\end{equation}
The numerical coefficients then reflect different 'volumes'
of the halo, core and background in the space of proper elements $(a_{\rm p}, e_{\rm p}, \sin I_{\rm p})$,
as defined in Table~\ref{core_halo_background}.

\begin{figure}
\centering
\includegraphics[width=8cm]{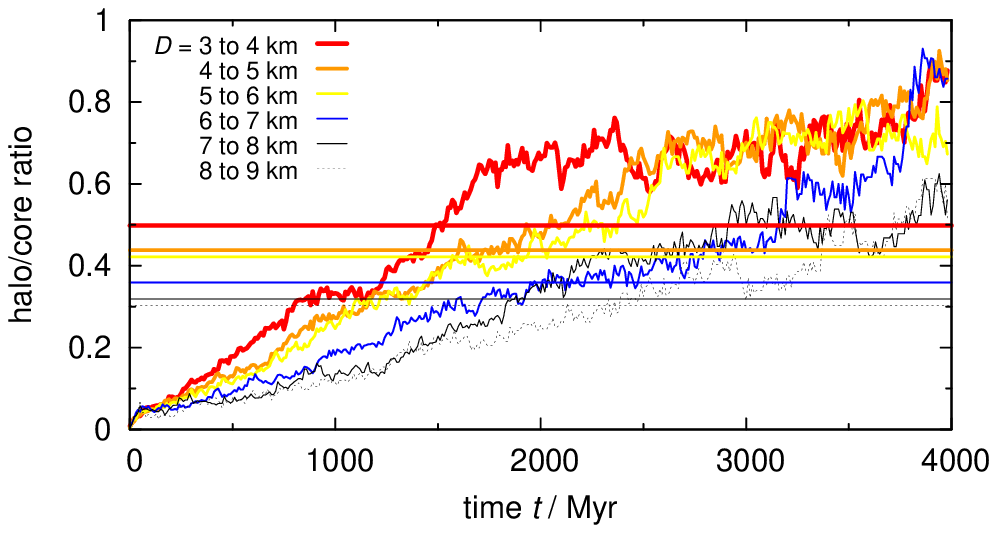}
\includegraphics[width=8cm]{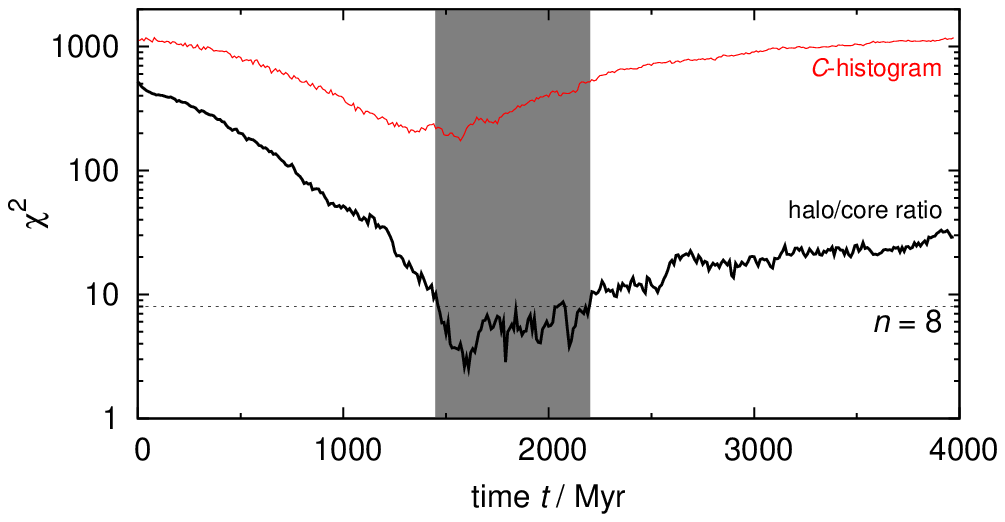}
\caption{Top panel: the evolution of the halo/core ratio in our simulation
for several size bins (colour curves) and its comparison to the observed SDSS ratios
in the same bins (horizontal lines). The intersections give age estimates
from 1.5\,Gyr (for smaller bodies) to 2.2\,Gyr (larger bodies). 
Bottom panel: the corresponding evolution of the~$\chi^2$ vs time~$t$.
The dotted line indicates the number~$n = 8$ of size bins (from $D = 2\,{\rm km}$ to $10\,{\rm km}$)
in which the $\chi^2$ was computed. The best fits (with $\chi^2 \simeq n$ or smaller)
correspond to ages from 1.5 to 2.2\,Gyr.
For comparison, we also plot a $\chi^2(t)$ dependence (red line),
computed from the histogram of $C \equiv (a-3.019\,{\rm AU}) / 10^H$ values,
which corresponds to the analysis of \citet{Vokrouhlicky_etal_2006Icar..182...92V}.
We use the population of Eos-like asteroids observed by the SDSS
in both core and halo for this purpose.}
\label{eos-5_halo_core_RATIO_BACKGROUND_chi2_CHIST}
\end{figure}

As we can see in Figure~\ref{eos-5_halo_core_RATIO_BACKGROUND_chi2_CHIST},
a reasonable match to the observed halo/core ratios can be obtained
for ages 1.5\,Gyr (for smaller bodies) to 2.2\,Gyr (for larger bodies).
To better quantify the difference between the model and the observations
we construct a suitable metric
\begin{equation}
\chi^2(t) \equiv \sum_{i=2}^9 { ({\cal R}_i(t) - {\cal R}_{{\rm obs}i})^2 \over \sigma_i^2(t) + \sigma_{{\rm obs}i}^2 }\,,\label{chi2}
\end{equation}
where the summation is over the respective size bins $(D_i, D_i+{\rm d} D)$,
$D_i \equiv i \cdot 1\,{\rm km}$ and ${\rm d} D = 1\,{\rm km}$.
The uncertainties of the numbers of objects are of the order
$\sigma_{\rm halo} \simeq \sqrt{{\rm d}N_{\rm halo}}$,
$\sigma_{\rm core} \simeq \sqrt{{\rm d}N_{\rm core}}$,
and $\sigma_i$ reflects their propagation
during the calculation of the ratio in Eq.~(\ref{ratio}) in a standard way
$$\sigma_i = \sqrt{ (\sigma_{\rm halo}/{\rm d}N_{\rm halo})^2 + (\sigma_{\rm core}/{\rm d}N_{\rm core})^2 } \, {{\rm d}N_{\rm halo}/{\rm d}N_{\rm core}}$$
and similarly for $\sigma_{{\rm obs}i}$.
The $\chi^2(t)$ dependence is shown in Figure~\ref{eos-5_halo_core_RATIO_BACKGROUND_chi2_CHIST}
and the best-fit is obtained again for the ages $t \simeq 1.5\hbox{ to }2.2\,{\rm Gyr}$.

The ratios~${\cal R}$ are directly related to the size-frequency distributions
and consequently we are indeed able to match the observed SFD's of halo and core,
including their slopes and absolute numbers (Figure~\ref{eos-5_halo_core_ae_0000My}, right column)

These results are {\em not\/} very sensitive to the initial velocity field,
because most asteroids fall within the family core;
velocities would be unreasonably large to have a substantial halo population initially.

%%%%%%%%%%%%%%%%%%%%%%%%%%%%%%%%%%%%%%%%%%%%%%%%%%%%%%%%%%%%%%%%%%%%%%%%

\section{Conclusions}\label{sec:conclusions}

Yarkovsky-driven origin seems to be a natural explanation of the halo population.
A lucky coincidence that the disruption of the Eos-family parent body
occurred close to the moderately strong 9/4 mean motion resonance with Jupiter
established a mechanism, in which orbits drifting in semimajor axis due to the
Yarkovsky effect are mostly perturbed by this resonance and scattered around
in eccentricity and inclination.
The total spread of the simulated halo (up to 0.2 in eccentricity, Figure~\ref{eos-5_halo_core_ae_0000My}),
which matches the SDSS observations (Figure~\ref{eos_halo_morby_BOX_ei_COLOR}), also supports our conclusion.

As an important by-product, the process enabled us to independently
constrain the age of the family.
Moreover, if we analyse the evolution in the proper semimajor axis vs the absolute magnitude $(a_{\rm p}, H)$ plane
and create a histogram of the quantity $C \equiv (a-3.019\,{\rm AU}) / 10^H$
(i.e. a similar approach as in \citet{Vokrouhlicky_etal_2006Icar..182...92V},
but now using a full $N$-body model and the SDSS observations for both the core and halo),
we can compute an independent $\chi^2(t)$ evolution
(refer to Figure~\ref{eos-5_halo_core_RATIO_BACKGROUND_chi2_CHIST}, red line).
Since both methods -- the halo/core ratios ${\cal R}$ and the $C$-histogram -- seem to be reasonable,
we can infer the most probable age as an overlap of intervals of low $\chi^2(t)$
and this way further decrease its uncertainty, so that $t \simeq 1.5\hbox{ to }1.9\,{\rm Gyr}$.

It is also interesting that the true anomaly at the time of disruption has to be
$f \simeq 120^\circ\hbox{ to }180^\circ$. We performed tests with lower values of~$f$
and in these cases the synthetic family has initially a different orientation
in the $(a_{\rm p}, e_{\rm p})$~plane: the objects are spread from small~$a_{\rm p}$ and~$e_{\rm p}$
to large~$a_{\rm p}$ and~$e_{\rm p}$ (cf. Figure~\ref{eos-5_halo_core_ae_0000My}).
Way too many objects thus initially fall in to the $z_1$~secular resonance
and because such captured orbits cannot drift to small~$a_{\rm p}$ and~large~$e_{\rm p}$
it is then impossible to explain the observed structure of the family
and consequently $f \lesssim 120^\circ$ is excluded.

Finally, let us emphasize that given the differences between the size-frequency distribution of the halo
that of the core, we can exclude a possibility that the Eos halo was created by a {\em purely\/}
gravitational process (like the perturbations arising from giant-planet migration).

%%%%%%%%%%%%%%%%%%%%%%%%%%%%%%%%%%%%%%%%%%%%%%%%%%%%%%%%%%%%%%%%%%%%%%%%

\section*{Acknowledgements}\label{sec:acknowledgements}

The work of MB has been supported by the Grant Agency of the Czech
Republic (grant no.\ 13-01308S)
and the Research Program MSM0021620860 of the Czech Ministry of Education.
We also thank both referees A. Cellino and V. Carruba
for careful reviews of this paper.

%%%%%%%%%%%%%%%%%%%%%%%%%%%%%%%%%%%%%%%%%%%%%%%%%%%%%%%%%%%%%%%%%%%%%%%%

\bibliographystyle{elsarticle-harv}
\bibliography{references}

\end{document}